%Paper: hep-lat/9301017
%From: Gyan Bhanot <gyan@Think.COM>
%Date: Mon, 25 Jan 93 18:15:17 EST

% THERE ARE 8 POSTSCRIPT FILES AT THE END OF THE MANUSCRIPT. THESE ARE
% SEPARATED FROM THE Tex FILE AND EACH OTHER BY A BUNCH OF %%%%% SYMBOLS
% AND A MESSAGE SAYING WHICH FIGURE IT IS. CUT OFF THIS LINE AND
% MAKE A file.ps FOR EACH FIGURE WITH %! AS THE FIRST LINE IN EACH FILE.
% NOW A SIMPLE lpr file.ps FROM ANY UNIX BOX WILL PRINT THE FILE.

%%%%%% BEGINNING OF TEX FILE

\magnification=1200
\tolerance=10000
\newcount\eqnumber
\eqnumber=1
\def\script{\scriptstyle}
\def\eqnam#1{\xdef#1{\the\eqnumber}}
\def\new{\the\eqnumber\global\advance\eqnumber by 1}
\footline={\ifnum\pageno>1\hfil\folio\hfil\else\hfil\fi}
\pageno=1
\rightline{IASSNS-HEP-93/1}
\rightline{WU B 93-05}
\rightline{Jan 1993}
\vskip.6in
\centerline{\bf Low Temperature Expansions for Potts Models}
\bigskip
\medskip
\centerline{\it by}
\bigskip
\medskip
\centerline{\bf Gyan Bhanot$^{a,b}$, Michael Creutz$^c$}
\centerline{\bf Uwe Gl\"assner$^e$, Ivan Horvath$^{c,d}$}
\centerline{\bf Jan Lacki$^b$, Klaus Schilling$^e$,
                John Weckel$^f$}
\bigskip
\bigskip
\baselineskip=18pt

{\it
\vfootnote{$^a$}{Thinking Machines Corporation, 245 First Street,
               Cambridge, MA 02142, USA}
\vfootnote{$^b$}{Institute for Advanced Study, Princeton, NJ 08540, USA}
\vfootnote{$^c$}{Brookhaven National Laboratory, Upton, NY 11973, USA}
\vfootnote{$^d$}{Physics Department, University of Rochester, Rochester,
                 NY, 14627, USA}
\vfootnote{$^e$}{Physics Department, University of Wuppertal, Gaussstrasse 20}
\vfootnote{ }{D-W-5600, Wuppertal 1, Germany}
\vfootnote{$^f$}{Physics Department, Princeton University, Princeton,
                 NJ, 08544, USA} } {\narrower \par On simple cubic
lattices, we compute low temperature series expansions for the energy,
magnetization and susceptibility of the three-state Potts model in $D=2$
and $D=3$ to $45$ and $39$ excited bonds respectively, and the
eight-state Potts model in $D=2$ to $25$ excited bonds.  We use
a recursive procedure which enumerates states explicitly.
We analyze the series using Dlog Pade analysis and
inhomogeneous differential approximants.

\vfill
\eject

A subset of the present authors recently described a method [1,2]
similar to the finite lattice method [3] for generating low
temperature series for discrete models. This method is based on a
recursive computer enumeration of configurations and has resulted in
series expansions for the $D=3$ Ising model that extend
available series by several terms [2-4].

In this paper, we present results from a similar analysis for the low
temperature expansions of Potts models in two and three
dimensions on a simple cubic lattice. We will not describe the method
used in much detail. It has already been outlined in Ref~[2] and will
be described in detail in a separate paper.

The energy for the $q$-state Potts model is defined to be
$$
E=\sum_{\langle ij \rangle}\left[1-\delta_{\sigma _i,\sigma _j}
\right]  \eqno(\new)
$$
where $\sigma _i$ is a site-defined field that takes $q$ possible
values. The sum is taken over all nearest neighbor pairs of spins
with $\delta$ being the Kronecker symbol.

The partition function is the sum of the Boltzmann weights over all
configurations
$$
Z=\sum_{\{\sigma\}} e^{-\beta E} \eqno(\new)
$$
Sorting configurations by energy, we rewrite this as a sum over $E$.
Defining $P(E)$ to be the number of states with a given energy $E$,
we have
$$
Z=\sum_{E=0}^{dN}P(E) u^{E}\eqno(\new)
$$
where $d$ is the number of dimensions, $N$ is the number of sites and
$u=e^{-\beta}$.

We compute the coefficients $P(E)$ exactly on small systems by
recursively assembling the system one site at a time.  The method
enables us to build up a lattice with arbitrary length in one
direction.  Intermediate stages require an explicit enumeration of
exposed slices transverse to this direction.  This effectively
reduces the computational complexity to that of a system of one less
dimension.

The starting point is a list of all states and corresponding energies
for a single transverse layer of the lattice.  In $D=2$, the
transverse layer is a line of spins, in $D=3$, it is a
plane of spins.  All the spins outside this transverse layer
are frozen to the same value; that is, the boundary conditions in the
longitudinal direction are cold.  Spins are then sequentially freed to
build up the lattice in the longitudinal direction.  We store the
number of states with a given energy, $E$, and exposed top layer in an
array $p_0(E,I)$, where the integer $I$ is an index which specifies
the exact configuration of the exposed transverse layer using
bit-coding.  When a new spin is added, we obtain the new counts
$p_0^\prime(E,I)$ as a sum over the old counts $$
p_0^\prime(E,I)=\sum_{I^\prime} p_0(E-\Delta(I,I^\prime),I^\prime).
\eqno(\new)
$$ Here $I^\prime$ can differ from $I$ only in the bits representing
the newly covered spins, and $\Delta(I,I^\prime)$ is the change in
energy due to any newly changed bonds.  For the present analysis we add
the spins one at a time.  Thus, the sum in the above equation is only
over $q$ terms, representing the $q$ possible values of the newly
covered spin. After the lattice is grown, a sum over the top layers
gives the resulting ~$P(E)=\sum_I p_0(E,I)$.  We always continue this
recursion sufficiently to avoid finite size errors in the longitudinal
direction.

As the temperature goes to zero, so does the variable $u$.
Thus, what we have in Eq.~(3) is the low temperature expansion for $Z$.
{}From it, we compute the series for the average energy,
$\langle E \rangle =\left(u{\partial \over \partial u}\right)
\log(Z).$
Subtracting this expectation value before adding the last spin from
its value after adding the last spin, we obtain the average energy per new
site.
This also eliminates the effect of the fixed end boundaries.
Writing,
$$
\langle E/N \rangle = \sum_j e_j u^j\eqno(\new)
$$
the low temperature expansion amounts to listing the coefficients
$e_j$.

The recursive technique can be extended to enable calculation of
quantities such as the magnetization and susceptibility.  We define a
magnetization in the Potts model by
$$
\langle M \rangle=\sum_i\langle \delta_{\sigma_i,0}\rangle
= N \sum_j m_j u^j\eqno(\new) $$ assigning to each unexcited spin the
value one, and to each excited spin the value zero. The calculation of
susceptibility is carried out using the fluctuation-dissipation
theorem and we define the low temperature series coefficients $\chi
_j$ as follows: $$ N\chi = \langle M^2 \rangle - \langle M \rangle^2 =
N\sum_j \chi _j u^j \eqno(\new) $$ Let $p(E,M,I)$ to be the number of
states with given energy, magnetization and exposed top layer $I$. To
compute any moment of the magnetization, it would be sufficient to
compute $p(E,M,I)$.  However, one can avoid computing this
memory expensive quantity.  Let us demonstrate this for the case
of the magnetization.

First, note that ${p_0}(E,I) = \sum_M p(E,M,I)$ is the count we had
before. To compute the magnetization, we need one more count:
${p_1}(E,I)=\sum_M Mp(E,M,I)$. This is because the
expectation value of magnetization can be written as,
$$ \langle M
\rangle = {{\sum_E P_1(E)u^E}\over Z},\eqno(\new)
$$
with
${P_1}(E)=\sum_I{p_1}(E,I)$ and $Z=\sum_{E,I}{p_0}(E,I)u^E$.
The counting scheme for ${p_1}(E,I)$ is
easy to derive. In analogy with Eq.~(4) one can write,
$$\eqalign{
{p_1^\prime}(E,I)&=\sum_M Mp^\prime(E,M,I)=\sum_{M,I^\prime}
Mp(E-
       \Delta_e, M-\Delta_m,I^\prime)= \cr
      &=\sum_{M,I^\prime}(M-\Delta_m+\Delta_m)\,p(E-\Delta_e,
       M-\Delta_m,I^\prime) = \cr
      &= \sum_{I^\prime}\left[\,{p_1}(E-\Delta_e,I^\prime)
       + \Delta_m p_0(E-\Delta_e,I^\prime)\,\right]}\eqno(\new)
$$

Here $\Delta_e\equiv\Delta_e(I,I^\prime)$ and $\Delta_m\equiv
\Delta_m(I,I^\prime)$ denote the change in energy and magnetization
when adding the new spin. Thus, computation of the magnetization
series requires just the introduction of one additional count, (which
only doubles the memory requirement) and we can calculate the
magnetization series to essentially the same order as the energy
series.

For the susceptibility series, we need to compute
$\langle M^2 \rangle$. This requires a count
${p_2}(E,I)=\sum_M M^2p(E,M,I)$. It is easy to see that $p_2$
obeys the recursion relation,
$$
{p_2^\prime}(E,I)
       = \sum_{I^\prime}\left[\,{p_2}(E-\Delta_e,I^\prime)
       + 2{\Delta_m}p_1(E-\Delta_e,I^\prime)
       + {\Delta_m}^{2} p_0(E-\Delta_e,I^\prime)\,\right]\eqno(\new)
$$

As discussed in Ref~[2], we work on generalized helical lattices and
label our lattice points by their ordinal number on a helix.  In three
dimensions, the nearest neighbors on the lattice in the $x$, $y$ and
$z$ directions are separated by $h_x$, $h_y$ and $h_z$ steps along the
helix respectively.  We assume that the $h$'s are ordered so that $h_x
< h_y < h_z$. Then, our numerical method requires us to keep track of
at most $q^{h_z}$ states and so we try to make $h_z$ as small as
possible.  Let $n$ be the effective lattice size, defined as the
length of the shortest closed path on the helical lattice. For a given
set of $h$ values, if we compute the set of nonzero vectors
$S=\{n_x,n_y,n_z; n_xh_x+n_yh_y+n_zh_z=0\}$ then $n={\rm
Min}_S(|n_x|+|n_y|+|n_z|)$.  The series expansion will be correct up
to the order $u^{(4n-1)}$.  Higher orders are corrupted by
contributions from graphs that wrap around the lattice.  However, as
described in Ref~[2], we can combine results from different helical
lattices to cancel these finite size effects to some order in the
series.  In two dimensions, there is not enough complexity for this
cancellation mechanism to work. Instead, one observes that keeping
$h_y$ spins in the top layer, the optimal choice of the lattice is
$h_x=h_y-1$. This gives the series correct to order $4h_y-3$.

Our series are listed in Tables I-III. The series for $D=2$ and $D=3$
Potts Models were computed on a CM-200/CM-2 Connection Machine using
CM-Fortran and C* programs.  The $D=2$, 8-states model series were
computed on a CRAY-2 using a C code and checked on a CM-2 using
CM-Fortran code. To get 3d series up to $39$ excited bonds,
we used lattices of
effective size up to $10$. This required the top layer to have
at most 15 spins. In Table IV we show the lattices and combination
factors used.

Note that our definition of $M$ in Eq.~6 is such that in the
completely disordered state it has the value $N/q$. The proper
order parameter for Potts models is the so called reduced
magnetization $M_R$ which is related to $M$ by the formula
$M_R=[qM-N]/(q-1)$. The reduced magnetization takes the values N and 0
in the completely ordered and disordered states respectively.
The results we give below from our analysis of
series are for the reduced magnetization and the corresponding
susceptibility.

In addition to the usual dlog Pade (DlP) method [5,6], we will use the
method of inhomogeneous differential approximants (IDA) introduced by
Fisher and Au-Yang [7] (see also [8]). These are useful in handling
singularities of the form, $$ F(u)=A(u)(1-u/u_c)^\zeta +B(u)\eqno(\new) $$
where $A$ and $B$ are analytic in $u$.

Given a series expansion for $F(u)$ to $N$-th order,
$F_N(u)=1+\sum_{i=1}^N f_iu^i$, (we will use the simplification that one
can always normalize the series so that the constant term is unity),
one computes coefficients for polynomials $Q_L(u)=\sum_{i=0}^L
q_iu^i$, $R_M(u)=1+\sum_{i=1}^M r_iu^i$ and $S_J(u)=\sum_{i=0}^J
s_iu^i$, which satisfy,
$$
F_N Q_L +S_J =F'_N R_M\eqno(\new)
$$
to order $N$, with $L+M+J=N-2$.
Note that for $S_J=0$ one gets the usual Dlog Pade ratio from
$Q_L/R_M$.  It is easy to see that potential critical points $u_c$ are
the zeros of $R_M$ and for each of these, the exponent $\zeta$ is
estimated as $\zeta=-Q_L(u_c)/R'_M(u_c)$.

Consider first the $D=2$ Potts models. Here, we know from self-duality
that the critical point is at $u_c=1/(\sqrt{q}+1)$.
For $q\le4$ the transition is continuous and
the critical exponents are known exactly (see [9] and references
therein). Models with $q>4$ undergo a first order phase transition.
Having results from both of the above categories available, our $D=2$ series
offer themselves as a good testing ground for series analysis methods.

Given the low temperature series, does one has enough information to
determine the nature of the transition, assuming that the critical
temperature is exactly known. In $D=2$, because of self duality, this
is easy if the series at hand has a sufficient number of terms.  To
illustrate this, we plot in Fig.~1a the energy as a function of $u$
from the low temperature series and its dual high temperature series
for $q=3$ and $q=8$. In Fig.~1b and Fig.~1c, we plot the latent heat
$L(n)$ derived using duality at the known critical point as a function
of the number of terms $n$ in the series. The fits of $L$ to a power
law in $1/n$ (Fig.~1b and 1c) convincingly demonstrate that the $q=3$
model has a second order transition while the $q=8$ model has a first
order transition with the latent heat equal to $1/2$ to 2 parts in a
thousand.

In general however, self duality is not available as a symmetry.  In
this case, one must rely on DlP and IDA analysis on the low
temperature series to determine the critical properties. Our arguments
below are similar in spirit to the discussion presented by Enting and
Guttmann [10].

If the system undergoes a second order phase transition, one expects
in general that the order parameter $M_R/N$ vanishes at the critical
point, approaching it with infinite slope. Estimates of the critical
temperature (poles) from DlP-s should then cluster well around the
exact value and estimates of the critical exponent $\beta$
(residues) should also be quite accurate. On the other hand, at a
first order transition, the magnetization is finite and nonzero and
its slope can be either finite or infinite.  In this case one would
expect the approximants to continue the curve beyond the critical
temperature along the so called pseudo-spinodal line [11].  This line
intersects the temperature axis at the point $u_S$ with corresponding
exponent $\beta_S$. Applying DlP-s in this case should then result in
a systematic overestimation of the critical temperature because it is
$u_S$ that the Pade is trying to fit.

In case of a first order transition with a divergent slope of the
magnetization as the transition is approached, DlP-s still tend to
overestimate the transition point because the finite value of the
magnetization is not modelled in the DlP-s (more detailed reasons can
be found in [12]). However, for this case, the IDA-s should treat the
situation better because they can account for a finite $<M/N>$ at the
critical point.  Thus, comparing the results of the two types of
approximants one might be able to determine the order of the
transition.

Applying DlP-s to the 45-term magnetization series of the 3-state
model in $D=2$ leads to a slight systematic underestimation of the critical
point. Taking into account seven most central approximants we got
$u_c=0.36595\pm0.00003$ which is to be compared with the exact
value $u_c=0.36602...$. The error here corresponds to the
scattering of values from the different DlP-s. In the
light of the above discussion, this suggests that the transition is
continuous. We estimate the
critical exponent $\beta=0.1084\pm0.0002$ by evaluating it at the known
critical point for this model. The error bar is of course meaningless as
it comes only from the error on the extrapolation and ignores the
systematic effects of the finiteness of the series. The value obtained is
about 2.5$\%$ below the exact result $\beta=1/9$.

In the 8-state model in $D=2$ on the other hand, DlP-s show a
critical point at $u_c=0.2628\pm0.0003$ which is substantially beyond
the true value $u_c=0.2612...$. This suggests a first order phase
transition. In Fig.~2 we plot $u_c$ versus $\beta$
for small values of J. The points for different $J$ lie fairly well
on a line with an obvious tendency to overestimate the critical point
again. This again establishes the first order nature of the transition.
The corresponding pseudo-exponent estimated from DlP-s has the
value $\beta_S=0.059\pm0.005$.

Similar ideas can be applied to the energy and specific heat series.  At a
first order phase transition there is a finite latent heat but the energy
curve can have either finite or infinite slope (specific heat) as that
point is approached.  DlP analysis of the $q=3$ specific heat series in $D=2$
shows a slight overestimate of the critical point, namely
$u_c=0.36626\pm0.00001$.  IDA-s on the other hand lead to a small
underestimate (see Fig.~3) giving an overall consistency with the second
order phase transition present. DlP-s average for critical exponent
$\alpha=0.412\pm0.001$ is rather poor when compared to the exact value
$\alpha=1/3$. This is probably due to the strong confluent singularity
present in this case [13].  The results of the IDA analysis is shown in
Fig.~3 where we plot $u_c$ versus $\alpha$ for various $J$ values from
$0-20$ with $L$ and $M$ chosen to be equal or differing by at most one (see
Eq.~12).  Notice that if we fit the data to a straight line and compute the
value of $\alpha$ at the exactly known critical point (vertical line in
Fig.~3), we get a result which differs from the exact value by about
$1\%$.

In the $q=8$ model the results from the specific heat series and
magnetization series are very consistent with each other. There is an
overestimate of the critical point by DlP-s ($u_S=0.2620\pm0.0001$) as
well as by IDA-s.  The averaged pseudo-exponent from DlP-s
is $\alpha_S=0.592\pm0.004$.

Finally, an analysis of the susceptibility series for the $q=3$ model
using the Dlog Pade and IDA analysis gave $\gamma=1.47\pm0.02$ by
extrapolating to the known critical point as was done above for
$\alpha$ and $\beta$. This is to be compared with the exact result
$\gamma=13/9 = 1.444...$.  For the $q=8$ model, we estimate
$u_S=0.2629\pm0.0009$, $\gamma_S=1.16\pm0.07$.

Let us now turn to the series for the $q=3$ Potts model in $D=3$
given in Table III. Theoretically, this is the most interesting
case of those considered in this paper, because of its connection to
the SU(3) lattice gauge theory in $D=4$ [14] and because of the lack of
any exact results. There was a good deal of confusion about the nature
of the transition in the past but by now the first order nature of
this transition seems to be well established [15].  Although the transition
temperature is not known exactly, there are very accurate Monte Carlo
estimates for it. For the purpose of our analysis we will assume that
the value $u_c=0.57659(1)$ estimated in Ref. [15] is the exact result.
We will do so because we found that neither the DlP nor the IDA
analyses can yield a more accurate value.

Consider first the magnetization series. In Fig.~4 we show the results
from central Dlog Pades. The data clusters well around the value
$u_c=0.5785\pm0.0003$, quite far from $0.57659$.  IDA-s show the same
tendency as can be seen in Fig.~5.  Here, the results from small $J$
fall very nicely on a straight line beyond the critical point which is
marked by a cross. These results support the conclusion that this
model has a first order phase transition in agreement with [10] and
Monte Carlo data [15].  The critical pseudo-exponent from
DlP-s has the value $\beta_S=0.204\pm0.002$ which agrees very
accurately with results of Miyashita et al. [16] who analyzed a
shorter series, and also with numerical simulations [17].

Next consider the specific heat series. Here one gets stable results from
many central Dlog Pades. Also, the IDA-s are quite stable for small $J$.
Fig.~6 shows the results of these analyses. The circles correspond to the
DlP-s and the other symbols are the results from the IDA-s for $J\le 4$.
There is no clear evidence for systematic overestimation of the critical
point by neither DlP-s nor IDA-s suggesting that the transition is weakly
first order in this variable. The straight lines in Fig.~6 are least
square fits to IDA-s and DlP-s.  Since the latent heat is small, one would
expect that these should intersect at the critical point where they are
both dominated by the singularity. Away from the critical point, the Dlog
Pade and the IDA-s treat the non-leading corrections differently and so the
results from them could be different.  Indeed the lines in Fig.~6 intersect
at $u_c=0.5766(2), \alpha=0.421(2)$. We have estimated the error on these
parameters from the errors in the fitted parameters for the straight lines.

Finally, we analyzed the $q=3$ susceptibility series in three
dimensions.  Here the combined data for DlP and IDA fall nicely on a
line. We estimate $\gamma=1.085\pm0.005$ by evaluating the fitted line at
$u_c=0.57659$.

Recently, Vohwinkel [18] has extended the shadow lattice method and
shown how one can obtain extremely high order low temperature
expansions. His series for the magnetization has several more terms than
ours and although he does not generate series for the other quantities
we measure in the present paper, we presume he can do so. A challenge
now is to see if the ideas of Ref. [18] can be incorporated into our
method.

\vskip .5in
{\bf\noindent Acknowledgements}
\vskip .2in

The research of GB was partly supported by U.S.~DOE Grant
DE-FG02-90ER40542 and the Ambrose Monell Foundation.  The research of
MC and IH was supported by U.S.~DOE Grant DE-AC02-76CH00016.  The
research of JL was partly supported by the Swiss National Scientific
Fund. We thank Thinking Machines Corporation and SCRI-Florida State
University for time on their Connection Machines and also NERSC at
Livermore for time on their CRAY-2.  The work of KS was supported in
part by Deutsche Forschungsgemeinschaft, grant Schi 257/1-4. We are
grateful to Professors Guttmann and Enting for discussions.  GB thanks
Professors Steven Adler and Michael E. Fisher for discussions and
comments on the manuscript.

\vfill\eject
{\bf \noindent References}
\vskip .2in
\item{1.}
M.~Creutz, Phys.~Rev.~B43 (1991) 10659.
\item{2.}
G.~Bhanot, M.~Creutz and J. Lacki, Phys.~Rev.~Let.~69 (1992), 1841.
\item{3.}
T. de Neef and I.G. Enting, J. Phys. A 10, (1977) 801;
I.G.~Enting, Aust.~J.~Phys.~31 (1978) 515; A.J.~Guttmann and
I.G.~Enting, Nucl.~Phys.~B (Proc.~Suppl.) 17 (1990) 328.
\item{4.}
I.G.~Enting, A.J.~Guttmann , In preparation.
\item{5.}
M. E. Fisher, Rocky Mtn. J. Math. 4 (1974) 181.
\item{6.}
G. A. Baker, `Essentials of Pad\'e Approximants', Academic Press, NY, 1975.
\item{7.}
M. E. Fisher and H. Au-Yang, J. Phys. A. Vol. 12 (1979) 1677.
\item{8.}
D. L. Hunter and G. A. Baker, Phys. Rev. B19 (1979) 3808.
\item{9.}
F.Y.~Wu, Rev. Mod. Phys., Vol.54, No.1, (1982) 235.
\item{10.}
A.J.~Guttmann and I.G.~Enting in Lattice 90, Capri, Nucl. Phys. B
(Proc. Suppl.) 17, (1990) 328.
\item{11.}
B. Chu, M. E. Fischer and F. J. Schoenes, Phys. Rev. 185 (1969) 219.
\item{12.}
A.J.~Guttmann, Asymptotic analysis of power series expansions, in:
Phase Transitions and Critical Phenomena, Vol. 13, eds. C.~Domb and
J.~Lebowitz (Academic, London, 1989).
A. I. Liu and M. E. Fisher, Physica A 156 (1989) 35.
\item{13.}
I.G.~Enting, J. Phys. A 13, (1980) L133; J.~Adler, I.G.~Enting and
V.~Privman, J. Phys. A 16 (1983) 1967.
\item{14.}
B.~Svetitsky and L.~Yaffe, Nucl. Phys. B210[FS6], (1982) 423; Phys.
Rev. D26, (1982) 962.
\item{15.}
R. V. Gavai, F. Karsch, B. Petersson, Nucl. Phys. B322 (1989) 738.
\item{16.}
S.~Miyashita, D.D.~Betts and C.J.~Elliot J. Phys. A 12, (1979) 1605.
\item{17.}
H.J.~Hermann, Z.~Physik B 35, 171-175~(1979).
\item{18.}
C.~Vohwinkel, " Yet Another Way to Obtain Low Temperature Expansions
for Discrete Spin Systems", DESY Preprint, Nov., 1992.

\vfill\eject
{\bf \noindent Figure Captions}
\vskip .2in
\item{Figure 1a.}  The average energy from the series expansions in
$D=2$ for $q=3$ and $q=8$. Duality was used to get the series
in the high temperature phase from the series in the low temperature
phase. The exactly known transition points are shown as vertical
lines.

\item{Figure 1b.}  The Latent heat $L(n)$ as a function of $n$ for $q=3$
in $D=2$. The solid line is a fit to a power law and demonstartes that
for $n=\infty$, the latent heat vanishes.

\item{Figure 1c.}  The Latent heat $L(n)$ as a function of $n$ for $q=8$
in $D=2$. The solid line is a fit to a power law plus a constant and
demonstartes that for $n=\infty$, the latent heat is about $1/2$.

\item{Figure 2.} $u_c$ versus the exponent $\beta$ from the magnetization
series for the $q=8$ model in $D=2$. The exact value of $u_c$ is the
vertical line.

\item{Figure 3.} $u_c$ versus the exponent $\alpha$ from the series for
the specific heat for $q=3$ in $D=2$ from IDA analysis.
The vertical line is the exact value of $u_c$.

\item{Figure 4.} $u_c$ versus the exponent $\beta$ from the magnetization
series for the $q=3$ model in $D=3$ using DlP-s.

\item{Figure 5.} $u_c$ versus the exponent $\beta$ from the magnetization
series for the $q=3$ model in $D=3$ using IDA-s with small $J$ values.  The
`exact' value of $u_c$ is marked with a plus and is a Monte Carlo result
from Ref.~15.

\item{Figure 6.} $u_c$ versus the exponent $\alpha$ from the series for
the specific heat for $q=3$ in $D=3$ from DlP and IDA analysis. The
transition point is accurately determined by the crossing of the lines
for DlP-s and IDA-s.

\vfill\eject
{\baselineskip=9pt
\font\magnifiedsevenrm=cmr7 at 7.5pt
\font\magnifiedsevenbf=cmbx7 at 7.5pt
\magnifiedsevenrm
\vbox
{\noindent
{\magnifiedsevenbf Table I}: The low temperature expansion coefficients
$e_i$, $m_i$ and $\chi _i$ for the energy, magnetization
and susceptibility series for the $q=3$ Potts model in $D=2$
on a simple cubic lattice.
\medskip
\settabs 4 \columns
\hrule
\smallskip
{\magnifiedsevenbf
\+ i & e$_i$ & m$_i$ & c$_i$\cr
}
\smallskip
\hrule
\smallskip
\+ 0 & 0 & 1 & 0\cr
\+ 1 & 0 & 0 & 0\cr
\+ 2 & 0 & 0 & 0\cr
\+ 3 & 0 & 0 & 0\cr
\+ 4 & 8 & -2 & 2\cr
\+ 5 & 0 & 0 & 0\cr
\+ 6 & 24 & -8 & 16\cr
\+ 7 & 28 & -8 & 16 \cr
\+ 8 & 32 & -24 & 100\cr
\+ 9 & 216 & -72 & 216 \cr
\+ 10 & 160 & -140 & 844\cr
\+ 11 & 660 & -320 & 1,552\cr
\+ 12 & 2,072 & -1,164 & 7,844\cr
\+ 13 & 1,664 & -1,560 & 12,112 \cr
\+ 14 & 11,760 & -7,044 & 60,268 \cr
\+ 15 & 17,700 & -13,000 & 118,944\cr
\+ 16 & 41,088 & -35,984 & 424,072\cr
\+ 17 & 156,468 & -101,736 & 1,081,392\cr
\+ 18 & 207,240 & -219,616 & 3,201,728\cr
\+ 19 & 849,300 & -647,536 & 8,670,688\cr
\+ 20 & 1,817,048 & -1,602,194 & 25,713,154\cr
\+ 21 & 4,021,780 & -3,970,384 & 67,206,560\cr
\+ 22 & 13,178,264 & -11,239,056 & 203,077,760 \cr
\+ 23 & 25,754,296 & -26,891,584 & 532,881,432 \cr
\+ 24 & 75,653,408 & -73,534,214 & 1,558,159,918\cr
\+ 25 & 193,458,400 & -191,374,464 & 4,250,639,632\cr
\+ 26 & 440,725,376 & -486,815,472 & 11,956,293,152\cr
\+ 27 & 1,296,485,460 & -1,323,802,480 & 33,296,697,848\cr
\+ 28 & 3,009,317,200 & -3,380,001,144 & 92,820,406,096 \cr
\+ 29 & 7,977,739,920 & -8,964,296,480 & 257,249,275,776\cr
\+ 30 & 21,217,637,824 & -23,766,809,488 & 721,023,458,656 \cr
\+ 31 & 51,359,965,976 & -61,628,612,552 & 1,986,080,278,600 \cr
\+ 32 & 140,885,970,816 & -165,028,619,666 & 5,561,045,323,298 \cr
\+ 33 & 354,038,121,756 & -432,231,505,864 & 15,359,165,767,512 \cr
\+ 34 & 916,153,258,448 & -1,142,608,252,368 & 42,717,426,328,784 \cr
\+ 35 & 2,439,917,838,708 & -3,039,729,276,192 & 118,457,421,095,792\cr
\+ 36 & 6,161,990,034,800 & -7,994,207,679,356 & 328,170,466,563,836\cr
\+ 37 & 16,397,314,674,708 & -21,295,402,476,752 & 909,829,346,983,664\cr
\+ 38 & 42,540,620,667,584 & -56,399,959,949,412 & 2,520,622,606,225,868 \cr
\+ 39 & 110,314,458,936,968 & -149,510,058,508,096 & 6,973,368,153,491,880\cr
\+ 40 & 292,427,669,006,272 & -398,341,255,729,746 & 19,322,697,243,220,158\cr
\+ 41 & 756,553,239,055,504 & -1,056,154,269,407,136 &
53,409,977,638,363,032\cr
\+ 42 & 1,994,873,374,110,312 & -2,813,530,068,950,904 & \cr
\+ 43 & 5,238,354,130,103,568 & -7,489,714,245,193,504& \cr
\+ 44 & 13,686,401,970,717,088 & -19,928,407,714,223,232 & \cr
\+ 45 & 36,195,015,152,016,276 & -53,175,417,534,052,136 & \cr
\smallskip
\hrule
}
}
\vfill\eject

{\baselineskip=10pt
\font\magnifiedsevenrm=cmr7 at 8pt
\font\magnifiedsevenbf=cmbx7 at 8pt
\magnifiedsevenrm
\vbox
{\noindent
{\magnifiedsevenbf Table II}: The low temperature expansion coefficients
$e_i$, $m_i$ and $\chi _i$ for the energy, magnetization
and susceptibility series for the $q=8$ Potts model in $D=2$
on a simple cubic lattice.
\medskip
\settabs 4 \columns
\hrule
\smallskip
{\magnifiedsevenbf
\+ i & e$_i$ & m$_i$ & c$_i$\cr
}
\smallskip
\hrule
\smallskip
\+ 0 & 0 & 1 & 0\cr
\+ 1 & 0 & 0 & 0\cr
\+ 2 & 0 & 0 & 0\cr
\+ 3 & 0 & 0 & 0\cr
\+ 4 & 28 & -7 & 7\cr
\+ 5 & 0 & 0 & 0\cr
\+ 6 & 84 & -28 & 56\cr
\+ 7 & 588 & -168 & 336\cr
\+ 8 & -588 & 91 & 0\cr
\+ 9 & 4,536 & -1,512 & 4,536\cr
\+ 10 & 11,760 & -4,060 & 14,504\cr
\+ 11 & -13,860 & 0 & 15,792\cr
\+ 12 & 205,072 & -68,859 & 288,169\cr
\+ 13 & 144,144 & -84,840 & 556,752\cr
\+ 14 & 271,460 & -256,424 & 2,062,088\cr
\+ 15 & 7,553,700 & -2,678,760 & 15,132,264\cr
\+ 16 & -713,692 & -2,049,229 & 25,582,802\cr
\+ 17 & 45,219,048 & -21,023,016 & 165,495,792\cr
\+ 18 & 232,853,880 & -93,466,856 & 720,185,368\cr
\+ 19 & -14,850,780 & -107,162,496 & 1,588,846,728\cr
\+ 20 & 2,822,644,748 & -1,187,630,969 & 10,588,862,669\cr
\+ 21 & 6,212,314,080 & -3,159,741,984 & 33,856,668,720\cr
\+ 22 & 8,166,041,884 & -7,756,117,236 & 108,773,186,200\cr
\+ 23 & 131,708,763,816 & -56,277,329,304 & 596,266,427,232\cr
\+ 24 & 167,481,870,528 & -118,516,443,339 & 1,709,093,729,238\cr
\+ 25 & 846,878,642,400 & -506,752,816,584 & 7,126,592,218,032\cr
\smallskip
\hrule
}
}
\vfill\eject

{\baselineskip=10pt
\font\magnifiedsevenrm=cmr7 at 8pt
\font\magnifiedsevenbf=cmbx7 at 8pt
\magnifiedsevenrm
\vbox
{\noindent
{\magnifiedsevenbf Table III}: The low temperature expansion coefficients
$e_i$, $m_i$ and $\chi _i$ for the energy, magnetization
and susceptibility series for the $q=3$ Potts model in $D=3$
on a simple cubic lattice.
\medskip
\settabs 4 \columns
\hrule
\smallskip
{\magnifiedsevenbf
\+ i & e$_i$ & m$_i$ & c$_i$\cr
}
\smallskip
\hrule
\smallskip
\+ 0 & 0 & 1 & 0\cr
\+ 1 & 0 & 0 & 0\cr
\+ 2 & 0 & 0 & 0\cr
\+ 3 & 0 & 0 & 0\cr
\+ 4 & 0 & 0 & 0\cr
\+ 5 & 0 & 0 & 0\cr
\+ 6 & 12 & -2 & 2\cr
\+ 7 & 0 & 0 & 0\cr
\+ 8 & 0 & 0 & 0\cr
\+ 9 & 0 & 0 & 0\cr
\+ 10 & 60 & -12 & 24\cr
\+ 11 & 66 & -12 & 24\cr
\+ 12 & -168 & 28 & -56\cr
\+ 13 & 0 & 0 & 0\cr
\+ 14 & 420 & -90 & 270\cr
\+ 15 & 900 & -180 & 540\cr
\+ 16 & -1,728 & 318 & -930\cr
\+ 17 & -2,448 & 432 & -1,296 \cr
\+ 18 & 6,708 & -1,320 & 4,768 \cr
\+ 19 & 9,462 & -1,992 &  7,968\cr
\+ 20 & -14,280 & 2,760 & -10,560\cr
\+ 21 & -49,686 & 9,368 & -36,992\cr
\+ 22 & 71,940 & -14,460 &  64,812\cr
\+ 23 & 177,192 & -35,280 &  163,440.\cr
\+ 24 & -194,544 & 36,680 & -16,5464 \cr
\+ 25 & -684,300 & 134,568 & -659,088 \cr
\+ 26 & 515,892 & -108,516 & 600,024\cr
\+ 27 & 3,087,234 & -609,692 & 3,278,256\cr
\+ 28 & -1,927,296 & 370,500 & -1,980,408\cr
\+ 29 & -10,943,904 & 2,153,016 & -12,285,816\cr
\+ 30 & 3,863,712 & -792,218 & 5,005,014\cr
\+ 31 & 44,383,506 & -8,867,580 & 55,200,864\cr
\+ 32 & -4,406,976 & 935,124 & -6,062,712 \cr
\+ 33 & -177,069,948 & 34,889,512 & -227,203,096 \cr
\+ 34 & -1,133,220 & 63,834 & 1,954,650\cr
\+ 35 & 652,560,090 & -130,265,472 & 914,339,736 \cr
\+ 36 & 199,263,288 & -39,322,372 & -\cr
\+ 37 & -2,553,456,210 & 507,892,056 & -\cr
\+ 38 & -1,235,636,652 & 239,776,590 & -\cr
\+ 39 & 9,742,992,324 & -1,940,344,524 & -\cr
\smallskip
\hrule
}
}

\vfill\eject

{\baselineskip=10pt
\font\magnifiedsevenrm=cmr7 at 8pt
\font\magnifiedsevenbf=cmbx7 at 8pt
\magnifiedsevenrm
\vbox
{\noindent
{\magnifiedsevenbf Table IV}: The lattices parameters and combination
factors that give the series accurate to $39$ excited bonds in $D=3$.
\medskip
\settabs 4 \columns
\hrule
\smallskip
{\magnifiedsevenbf
\+ $h_x$ & $h_y$ & $h_z$ & Coefficient\cr
}
\smallskip
\hrule
\smallskip
\+ 9 & 14 & 15 & 2\cr
\+ 11 &12&15&-1 \cr
\+ 9 & 11&15&-2\cr
\+ 10 &13&14&1\cr
\+ 11 &12&14&5\cr
\+ 9 & 11&14&-1\cr
\+ 7 & 12&13&1\cr
\+ 10 &11&13&-3\cr
\+ 8 & 10&13&1\cr
\+ 5 & 11&12&3\cr
\+ 7 & 10&12&-5\cr
\smallskip
\hrule
}
}

\vfill\eject

\bye